\documentclass[aps,reprint,superscriptaddress,noeprint,amsmath,amssymb,prb,showkeys,floatfix,nofootinbib]{revtex4-2}

\usepackage[dvipsnames]{xcolor}
\usepackage{graphicx}
\usepackage[hidelinks=true,colorlinks=true,linkcolor=blue,citecolor=blue]{hyperref}
\usepackage[utf8]{inputenc}
\usepackage[T1]{fontenc}
\usepackage{physics}
\usepackage{mathtools}
\usepackage{siunitx}
\usepackage{bm}
\usepackage{dcolumn}
\usepackage{appendix}
\usepackage{lipsum}
\usepackage{multirow}
\usepackage{placeins}
\newcommand{\orcid}[1]{\href{https://orcid.org/#1}{\includegraphics[width=8pt]{orcid.pdf}}}
\usepackage[normalem]{ulem}

\let\oldAA\AA
\renewcommand{\AA}{\text{\normalfont\oldAA}}

\begin{document}

\title{Machine-Learning Surrogate Model for Accelerating the Search of Stable Ternary Alloys}
\author{M. Minotakis}
\affiliation{School of Physics and CRANN Institute, Trinity College, Dublin 2, Ireland}
\author{H. Rossignol}
\affiliation{School of Physics and CRANN Institute, Trinity College, Dublin 2, Ireland}
\author{M. Cobelli}
\affiliation{School of Physics and CRANN Institute, Trinity College, Dublin 2, Ireland}
\author{S. Sanvito}
\affiliation{School of Physics and CRANN Institute, Trinity College, Dublin 2, Ireland}
\date{\today}

\begin{abstract}
The prediction of phase diagrams in the search for new phases is a complex and computationally intensive
task. Density functional theory provides, in many situations, the desired accuracy, but its throughput becomes
prohibitively limited as the number of species involved grows, even when used with local and semi-local 
functionals. Here, we explore the possibility of integrating machine-learning models in the workflow for the
construction of ternary convex hull diagrams. In particular, we train a set of spectral neighbour-analysis 
potentials (SNAPs) over readily available binary phases and we establish whether this is good enough to predict the energies of novel ternaries. Such a strategy does not require any new calculations specific for the construction of the model, but 
just avails of data stored in binary-phase-diagram repositories. We find that a so-constructed SNAP is capable
of accurate total-energy estimates for ternary phases close to the equilibrium geometry but, in general, is not
able to perform atomic relaxation. This is because during a typical relaxation path a given phase traverses 
regions in the parameter space poorly represented by the training set. Different metrics are then investigated
to assess how an unknown structure is well described by a given SNAP model, and we find that the standard
deviation of an ensemble of SNAPs provides a fast and non-specie-specific metric.
\end{abstract}

\maketitle

\section{Introduction}

The rapid advancements in existing technologies such as logic electronic devices, as well as the need to develop
new energy solutions, make the search for novel materials of paramount importance. Following the improvement 
in hardware performance and the development of user-friendly {\it{ab initio}} algorithms for materials 
modelling~\cite{VASP,QE,ABINIT}, it has become possible to use such computational methods to accelerate the 
discovery process. Successful examples of theory-driven materials discovery include the development of Li-ion 
cathodes \cite{cathodematerials}, high entropy alloys \cite{HEA} and magnetic materials \cite{magmat}. 
The state-of-the-art workflow takes the form of a high-throughput search~\cite{CurtaNM}. This makes use of 
efficient density-functional-theory (DFT) calculations to predict the material properties of a large pool of compounds 
created virtually, in the hope that these contain some `hidden compounds' yet to be discovered \cite{oganovbook}. 
Collections of such prototypes exist in material databases \cite{AFLOW,materialsproject}. The first step in the workflow
is a stability screening, which ensures that any compound selected is chemically reasonable and thermodynamically stable, namely that it might form.  
Such screening requires the calculation of the total energy of the system. In this work we focus on demonstrating that
machine learning (ML) can assist this first step.

Two key ingredients are needed to find new stable phases: a method for generating candidate structures and a method 
for evaluating their energies. The first aspect, which is not the focus of this work, is usually performed by either constructing 
a library of prototypes \cite{magmat} or via dedicated methodologies for proposing novel structures \cite{AIRSS,GeneticAlgo,GOFEE}. 
Total energies are then typically evaluated by local/semi-local DFT, which ensures a general good accuracy at a moderate
computational cost. Unfortunately, even when combined with simple approximations, performing \textit{ab initio} calculations 
remains the rate-limiting step of any materials search workflow, posing constraints on both the number of candidate structures 
that can be tested and the number of atoms in the unit cells of such structures. In order to speed up this process ML has been 
deployed in a number of ways. For instance, it has been used to predict better starting charge densities for the self-consistent 
loop needed by Kohn-Sham DFT~\cite{brockherdeML_density,ML_density,ML_Bruno}. 
Furthermore, ML has also been used in active-learning frameworks to accelerate geometry optimisation \cite{kitchin}, as well 
as in \textit{ab initio} molecular dynamics (AIMD) \cite{VASPotf} through the direct prediction of energies and forces. In these 
cases ML models are used to map local atomic configurations to energies and forces, by tuning parameters based on training 
data provided from DFT. Once the models are trained, subsequent predictions are orders of magnitude faster than DFT 
calculations. We follow this approach in our work, by developing a machine learning model acting as a surrogate for the 
energy predictions made by DFT.

Several approaches have been employed to construct energies surrogate to the DFT ones in the computational materials 
discovery process~\cite{clusterexpansionsemi,CEternaries}. A commonly used strategy is the cluster expansion, in which 
the energy of a system is written as a sum of energy contributions from different clusters made up by the constituent 
atoms. Here, the strengths of the effective cluster interactions are computed by fitting the model to a set of DFT energies. 
Thus, this surrogate model uses DFT data to perform interpolation on new structures \cite{clusterexpansion}, a task that
is also performed when using machine learning techniques. A swarm of studies have utilised this class of methods in the context 
of materials discovery. These include for instance component predictions \cite{componentpred}, where the most likely compositions 
for a given set of atomic species is computed, or structure predictions \cite{strucpred}, where the most likely crystal structure 
is forecast, as well as for the direct prediction of the distance of a material's energy from the convex hull \cite{chulldist1,chulldist2}. 
In general, these schemes make use of databases with a variety of materials made of different species, in combination with 
various machine learning algorithms such as Bayesian optimisers, Random Forests and Support Vector Machines. The feature 
vectors typically include general information about the material structure as well as species characteristics of the constituent 
atoms. However, they are not typically trained to distinguish small differences between compounds with similar structures. 
For this task one has to rely on Machine Learning Interatomic Potentials (MLIAPs). 

MLIAPs combine fingerprints of the atomic configurations, most frequently locally defined, with a ML algorithm to predict energies, 
forces and stress tensors. Several atomic fingerprints, acting as the feature vectors, have been successfully deployed. Their locality
enforces invariance under translation and atomic permutation, while they are usually constructed to be locally rotational invariant.
The most successful MLIAPs notably include Behler-Parinello symmetry functions, combined with Neural Networks in Neural Network 
Potentials (NNP) \cite{behler2007generalized}, bispectrum components, combined with ridge regression in Spectral Neighbour Analysis 
Potential (SNAP)~\cite{SNAP} and quadratic SNAP (qSNAP) \cite{qSNAP}, the Smooth Overlap of Atomic Position (SOAP) descriptors with Gaussian process regression 
in Gaussian Approximation Potentials (GAP) \cite{bartok2010gaussian}, invariant polynomials with linear regression in Moment Tensor 
Potentials (MTP) \cite{MTP} and $N$-bond basis functions with linear regression in Atomic Cluster Expansion (ACE) \cite{ACE}. 
As for the cluster expansion method, MLIAPs make use of a DFT dataset to fit the model parameters and are capable of predicting 
energies and forces at \textit{ab initio} accuracy, provided these are made for structures for which the model 
interpolates \cite{MLIAPperformance}. This makes such potentials ideal for molecular dynamic simulations at high accuracy, for large 
systems and over long timescales \cite{SNAPalloys2,NNPCu,amorphousC,amorphousSi}.

In the context of predicting materials stability, MLIAPs are used to map the potential energy surface of multiple phases and hence 
to reconstruct the $T=0$~K phase diagram to determine the lowest energy structures, namely to construct the convex hull. SNAPs and 
NNPs have been previously used against this task for metallic alloys \cite{SNAPalloys1,NNPAuLi,NNPMgCa,NNPCuPdAg}. In the case 
of SNAP, however, the range of different structures and stoichiometries probed was limited and the potential was not used to find new 
stable alloys. Instead, in the case of NNPs, the training set used was very large ($\sim$ 10$^{3}$ - 10$^{4}$ structures), so that the
actual structures computed by DFT were as many as those needed to construct a fully \textit{ab initio} convex hull. This is also the 
case of the GAP models trained as general potentials across the phase diagram of C \cite{GAPGeneralC} and Si \cite{GAPGeneralSi}. 
One of the very few examples of using a MLIAP trained over a limited number of structures to predict materials stability at an accelerated
pace has been recently provided by Gubaev \textit{et al.} \cite{gubaev2019}. In their work, \textit{ab inito} calculations were performed 
over between 383 and 976 structures to train a MTP able to reproduce binary and ternary convex hulls. These structures were selected 
through an efficient active-learning process \cite{shapeevactive}, which probed about 10$^{5}$ configurations for each phase 
diagram. This method proved that MLIAPs could accelerate the computational high-throughput search of new alloys. 

In this work we show how SNAPs can be used to drastically accelerate the search for new stable ternary intermetallic compounds,
without the need to generate large training DFT datasets. Our approach for constructing a MLIAP for rapid screening is similar in philosophy 
to the \textit{specialised} MLIAP training proposed by Artrith \textit{et al.} to compute the binary convex hull of Li$_{x}$Si \cite{LixSi}. 
Our selection of the training database, however, is different. In fact, rather than investing resources on curating a training set and 
on running DFT calculations for the sole purpose of training the MLIAP, we make use of existing materials convex hull databases, 
namely the AFLOWlib \cite{AFLOW} repository. In this way, data is already available and the computational efforts put into any 
\textit{ab initio} calculations are also relevant for the convex hull construction. In AFLOWlib, the binary phase diagrams are typically 
extensively explored, meaning that there is a significant range of data available and that there is a lower probability of finding new 
stable binary alloys. The ternary hulls are usually not as rich, despite there being a combinatorial explosion of the number of possible 
derivative structures that can be created from a prototype structure. This leaves more room for exploration. For ternary systems, the 
enthalpic term in the Gibbs free energy is still significant with respect to the entropic one, unlike quartenary compounds \cite{HvsSAFLOW}, 
meaning that free-energy calculations are still relevant. The method proposed here then exploits SNAP to guide the screening of the 
ternary space. 

The manuscript is organised as following. In the next section we will present the main computational ingredients needed for our
workflow, namely the AFLOWlib dataset, the DFT numerical implementation used and the SNAP model trained. Then we will proceed
with presenting the results for three prototypical ternary systems, namely one composed by noble metals Cu-Ag-Au, and two
composed by early, mid and late transition metals, namely Ti-Mo-Pt and Cd-Hf-Rh. In that section we will discuss the SNAP 
training and its performance against known ternary compounds and newly constructed prototypes. Finally, we will conclude.

\section{Methods}

This work explores the ability of SNAP to be used as an efficient energy predictor of novel 
ternary compounds. Our general philosophy, however, is to achieve such goal without generating
DFT data serving the sole purpose of training the ML model, but instead we aim at re-using
the same DFT data computed to construct the binary phase diagrams. As such, the constructed
ML model will have negligible computational payload. With this in mind the structures used
to train SNAP are all taken from the CHULL AFLOWlib database \cite{AFLOW-CHULL}. Then the
SNAP is trained on the total energies of binary crystalline compounds and tested on ternary 
materials, either in their equilibrium geometry or as suggested from a prototypes' generator.
Different error metrics that can be used to identify structures far from the training set are 
then assessed. Here, all DFT energy values are obtained with the VASP package \cite{VASP}. 
Computational details are given in the sections below.

\subsection{AFLOWlib data}

The training data for all the models are directly taken from the AFLOWlib repository~\cite{AFLOW} through the AFLOW-CHULL 
toolset \cite{AFLOW-CHULL}. The access and manipulation of the data is performed with the AFLOW application programming 
interface (API) \cite{TAYLOR2014178}. In particular, we consider three ternary systems, Cu-Ag-Au, Ti-Mo-Pt and Cd-Hf-Rh, for 
which we have extracted data for all the binary and ternary phases contained in AFLOWlib.

\subsection{DFT}
\label{sec:DFT}

\textit{Ab-initio} total-energy calculations are performed using the Vienna Ab initio Softwate Package~\cite{VASP} (VASP), which 
makes use of periodic boundary conditions, a plane wave basis set and the Projected-Augmented-Wave (PAW) method with 
pseudopotentials. All calculations are performed with a plane wave cutoff of 600~eV and an energy convergence criterion of 
10$^{-4}$~eV. The standard generalised gradient approximation as parameterised by Perdew, Burke and Ernzerhof 
\cite{PhysRevLett.77.3865,PhysRevLett.78.1396} is used throughout, together with the corresponding VASP pseudopotential 
library. 

For all DFT calculations we use the convergence criteria set of the AFLOW standard \cite{CALDERON2015233}. The $k$-mesh 
is constructed with the Monkhorst-Pack scheme and ensuring that the mesh is Gamma-centered for the hexagonal (\textit{hP}) 
and rhombohedral (\textit{hR}) Bravais lattices. The number of sampling points, $N_{i}$, is proportional to the norm of each 
corresponding reciprocal Bravais lattice vector, $\vec{b_{i}}$, and are minimised ensuring the following condition
\begin{equation}
N_\mathrm{KPPRA} \leq \textup{min} \left [ \prod_{i=1}^{3}N_{i} \right ] \times N.
\label{ineq:nkppra}
\end{equation}
Here, $N_\mathrm{KPPRA}$ is the number of $k$-points per reciprocal atom and $N$ the number of atoms in the cell. In particular,
$N_\mathrm{KPPRA}$ is chosen at 10,000 for all static calculations and at 6,000 for all geometry relaxations. The geometry relaxations
are considered converged when the atomic forces are smaller than $10^{-3}$~eV/\AA.

\subsection{Spectral Neighbour Analysis Potential}

Our machine-learning model of choice for total energy predictions is SNAP~\cite{SNAP}, which is here rapidly recalled. Like many 
MLIAPs, SNAP assumes that the total energy of a $N$-atom system (molecule or crystal) can be broken down into individual 
contributions, $E_{i}$, associated to each atom $i$ and element $Z_{i}$
\begin{equation}
E_\mathrm{tot} = \sum_{i}E_{i}^{Z_{i}}\:.
\label{eq:energy_contrib}
\end{equation}
Each energy contribution $E_{i}^{Z_{i}}$ is assumed to be linearly dependent on a feature vector, $\vec{B}^{Z_{i}}_{i}$, describing the local 
atomic environment, where the coefficient of expansion, $\vec{\alpha}^{Z_{i}}$, represent the training parameters of the model,
\begin{equation}
E^{Z_{i}}_{i} = \vec{\alpha}^{Z_{i}} \cdot \vec{B}^{Z_{i}}_{i}.
\label{eq:energy_contrib}
\end{equation}
In SNAP the descriptors, $\vec{B}^{Z_{i}}_{i}$, are the bispectrum components \cite{bartok2010gaussian}, quantities that are invariant
upon local rotations. 

In a nutshell, the atomic-neighbour density function within a sphere of radius $R_{c}$ and centered at the $i$-th atom at $\vec{R}_{i}$ 
can be written in terms of a sum over the neighbours $j$ within the sphere,
\begin{equation}
    \rho_{i}\left ( \vec{r} \right ) = \delta \left ( \vec{r} - \vec{R}_{i} \right ) + \sum_{j} w_{Z_{j}}  \delta \left ( \vec{r} - \vec{R}_{j} \right ) f_{c}\left ( R_{ij} \right ),
\label{eq:density}
\end{equation}
where $R_{ij}=|\vec{R}_{i}-\vec{R}_{j}|$, $w_{Z_{j}}$ are weights associated with each atomic species (treated as hyperparameters) and $f_{c}$ 
is a cut-off function, as defined by Behler and Parinello \cite{behler2007generalized}. Such density is then projected onto the four-dimensional 
sphere of radius $r_{0}$ and expanded over hyperspherical harmonics $U^{J}_{m',m}\left ( \theta ,\phi ,\theta _{0} \right )$ as,
\begin{equation}
\rho_{i}\left ( \vec{r} \right ) = \sum_{J=0}^{\infty }\sum_{m,m'=-J}^{J}c_{m',m}^{J}U^{J}_{m',m}\left ( \theta ,\phi ,\theta _{0} \right )\:.
\label{eq:expansion}
\end{equation}
Here, $J$, $m$ and $m'$ are parameters distinguishing the individual hyperspherical harmonics and $c_{m',m}^{J}$ are the appropriate 
expansion coefficients. Details on the conversion from $\vec{r}$ to the four-dimensional polar angles can be found in Ref.~\cite{bartok2013representing}. 
In practice, the expansion is truncated at $J=J_\mathrm{max}$. The bispectrum components are then built as an appropriate tri-product of 
the expansion coefficients, 
\begin{align*}
B_{i}^{J,J_{1},J_{2}} &= \sum_{m_{1}',m_{1}=-J_{1}}^{J_{1}}c_{m'_{1},m_{1}}^{J_{1}}\sum_{m_{2}',m_{2}=-J_{2}}^{J_{2}}c_{m'_{2},m_{2}}^{J_{2}}\\
&\quad \times \sum_{m',m=-J}^{J} C_{mm_{1}m_{2}}^{J,J_{1},J_{2}}C_{m'm'_{1}m'_{2}}^{J,J_{1},J_{2}}\left ( c_{m',m}^{J} \right )^{*}\:,
\label{eq:bispectrum}
\end{align*}
where $C_{mm_{1}m_{2}}^{J,J_{1},J_{2}}$ and $C_{m'm'_{1}m'_{2}}^{J,J_{1},J_{2}}$ are the Clebsch-Gordan coefficients, which 
determine the coupling between the different values of $J$. The individual components with $J$ values between 0 and $J_\mathrm{max}$ are then 
collected to form the vector $\vec{B}_{i}$.

The calculation of the bispectrum coefficients is performed with the LAMMPS software \cite{LAMMPS}, while the energy fitting with an in 
house Python library that makes use of the SCIKIT-LEARN package \cite{scikit-learn}. The hyperparameters of the model are $J_\mathrm{max}$, 
$R_{c}$ and the set of weights {$w_{Z_{i}}$}, which are optimised manually. 

\section{Results and Discussion}

The first ternary system selected comprises three noble metals Cu, Ag and Au. These display less complexity in their chemical 
behaviour than other transition metals with partially filled $d$-shells, so that they represent a good playground to present our concept. 
The process chosen for identifying novel ternary prototypes involves the training a SNAP model on data {\it already} available and 
only related to binary phases. The so-constructed ML model is then used to predict the energy, and hence the stability, of a range 
of ternary compounds that, by definition, do not appear in the training set. 

\subsection{Fitting energies for binary phases}

As a first step, a series of SNAP models are trained individually over each of the three binary systems, namely Ag-Au, Cu-Ag and 
Cu-Au, for which AFLOWlib contains 261, 191 and 263 structures, respectively. The associated unary systems are included as well. 
Although we could have proceeded by using the AFLOWlib energies directly, for consistency we have here preferred to re-run static 
DFT calculations, according to the standards presented in Section~\ref{sec:DFT}, for all the unary and binary structures. Within 
each binary system, the dataset is split into a training and a cross-validation set, while the ternary phases form the test set. The 
split is 80\% training and 20\% cross-validation, as suggested by the learning curves shown in the Appendix. Monte Carlo cross-validation 
is the strategy of choice for splitting the datasets. Since there is significant structural diversity in the datasets and in order to avoid the 
effect that a small subset of structures may have on the training, an ensemble of 10 SNAP models is trained for each binary system. The average of the predictions made by each SNAP model of the ensemble is used as the final output. We call this ensemble of SNAP models {\it SNAPs}. 
The cross-validation set is then used to manually optimise the hyperparameters $J_\mathrm{max}$, $R_{c}$ and {$w_{Z_{i}}$} for 
each species, using a grid search procedure. Details on the optimal hyperparameters selection are given in the Appendix. 

\begin{table}[hbt]
\centering
\caption{Summary of the average errors over 10 SNAP models for the three binary systems. T = training,
CV = cross-validation}
\label{tab:errors_single_binaries}
\begin{tabular}{l|l|l|l}
\hline\hline
Error (meV/atom)	& Ag-Au & Cu-Ag & Cu-Au \\
\hline
MAE (T)           & 21.5	& 2.7  & 5.9\\
RMSE (T)           & 41.2	& 3.7  & 9.2\\
MAE (CV)	& 24.2	& 7.6  & 12.5\\
RMSE (CV)	& 43.0	& 14.2 & 25.1\\
\hline\hline
\end{tabular}
\end{table}

\begin{figure}[!h]
    \centering
    \includegraphics[width=8cm]{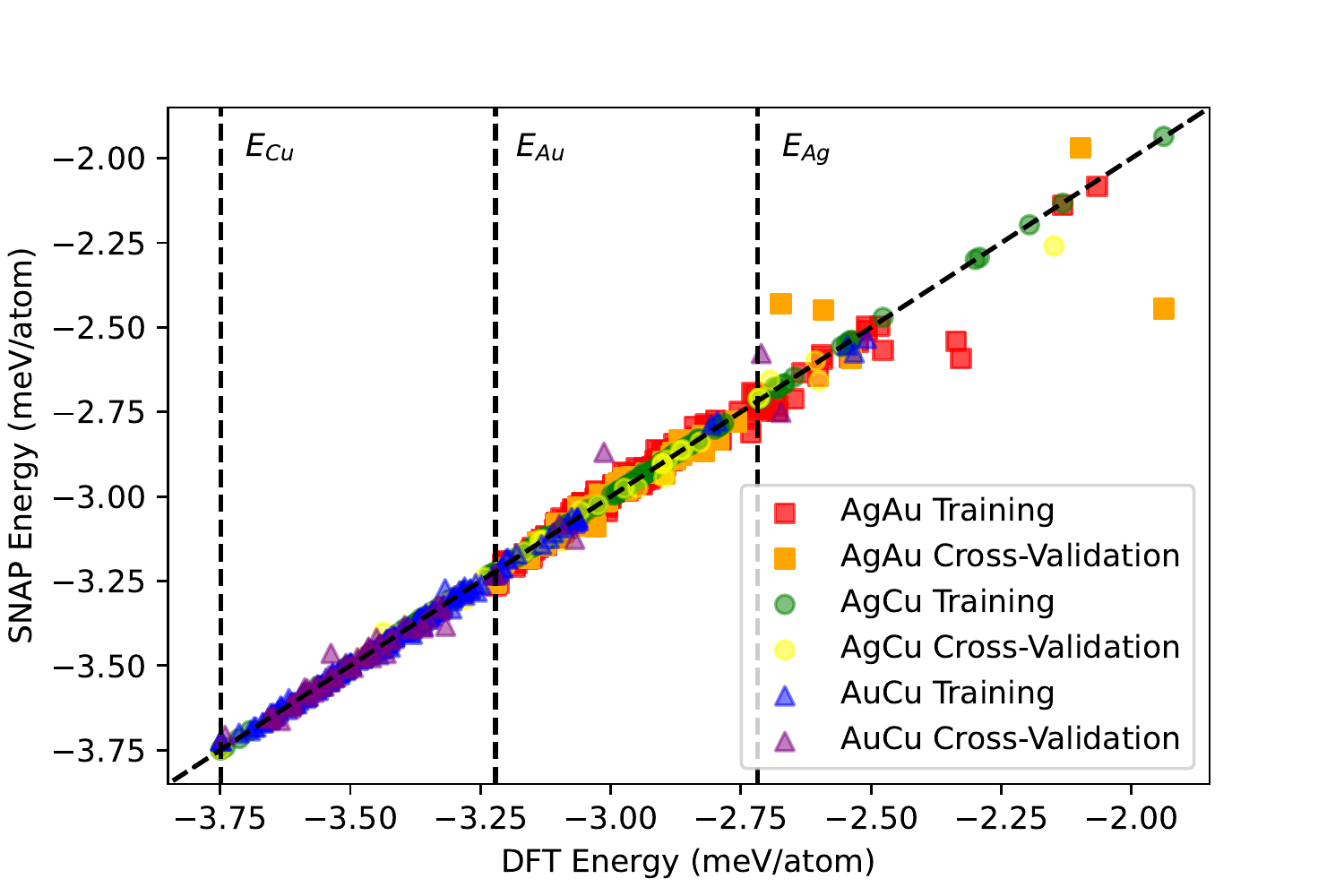}
    \caption{Parity plot, showing the SNAP-predicted energies against the DFT ones, for the Ag-Au, Cu-Ag and Cu-Au binary 
    systems. The DFT energies of the lowest-energy single-element structures are here marked by vertical dashed lines.}
    \label{fig:single_binary}
\end{figure}

The accuracy of our models is assessed by computing both the mean absolute error (MAE) and the root-mean squared 
error (RMSE), averaged over the 10 different SNAPs constructed for each binary system. The results are presented in 
Table~\ref{tab:errors_single_binaries} for both the training and the cross-validation set, while a visual appreciation of the 
performance is shown in 
the parity plot of Figure~\ref{fig:single_binary} that displays the results for one SNAP model randomly chosen from the ensemble. We find that the Cu-Ag system is better fitted, followed by Cu-Au, while
the error is larger for Ag-Au. This is attributed to the presence of a number of high-energy structures among the Ag-Au AFLOWlib
prototypes. These have been proved more difficult to fit, even when present in the training data, as demonstrated by their 
departure from the parity line in Figure \ref{fig:single_binary} (see red squares for SNAP energies larger than -2.75~meV/atom). 
When looking at the hyperparameter optimisation, we find the $R_{c}$ value of Ag-Au to be lower than that of the other
two binary systems. It is optimal to have similar weights for Ag and Cu, while distinguishing Au from the other 
two elements helps to reduce the error. 

\subsection{Testing over ternary compounds}\label{section:TestingTernaries}

\begin{figure}[!h]
    \centering
    \includegraphics[width=8cm]{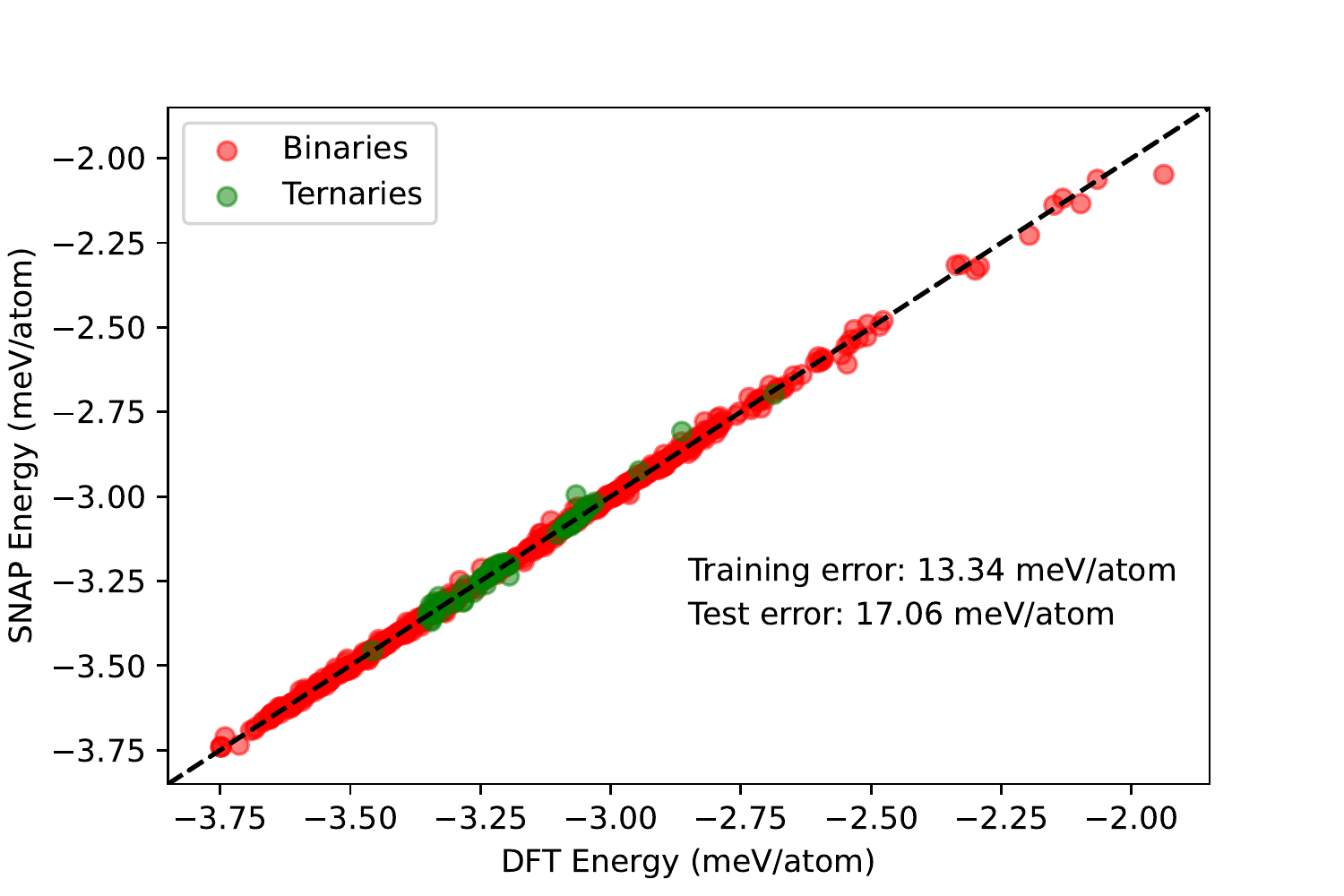}
    \caption{Parity plot, showing the SNAP-predicted energies against the DFT ones, for both the binary (training and cross-validation
    set) and ternary (test set) compounds contained in AFLOWlib for the Cu-Ag-Au system. Here the SNAP models forming the ensemble
    have been trained on the entire pool of binary phases. The RMSE is reported in the legend.}
    \label{fig:binaries_predict_ternaries}
\end{figure}

The SNAP models presented in the previous section have been individually trained over each binary system, namely
they contain information for only two species at a time. These cannot be utilized to predict a ternary structure. As such, 
we now train another ensemble SNAP, this time over the entire library of binary compounds available (261+191+263=715), by using
the same strategy described before (e.g. a 80/20 training/cross-validation split, see hyperparameters in the Appendix). Such
SNAP is then tested over the 78 ternary structures contained in AFLOWlib for Cu-Ag-Au. The parity plot for such a new model
is presented in Figure~\ref{fig:binaries_predict_ternaries}, where again data are presented for the SNAP prediction of on a model randomly chosen from
the 10. 

The figure clearly shows that our ensemble SNAP performs almost identically over the binary training set and the 
ternary test one, demonstrating that models having knowledge of enough binary structures are capable to offer
good energy predictions for ternaries. More quantitatively, the RMSE of the model is 13.34~meV/atom for the training
set, while it increases only marginally to 17.06~meV/atom for the test one. Interestingly, this is even lower than the
cross-validation error found for the binary individually trained SNAPs (see Table~\ref{tab:errors_single_binaries}),
a result that we attribute to the more extended diversity of the chemical environments that the model has now to fit.

It is therefore established that by training on data extracted from the binary phase diagrams SNAP models are able 
to predict the ground-state energy of fully relaxed ternary structures.    

\subsection{Ternary Prototypes}\label{sec:TernaryProto}

\begin{figure}[!h]
    \centering
    \includegraphics[width=8cm]{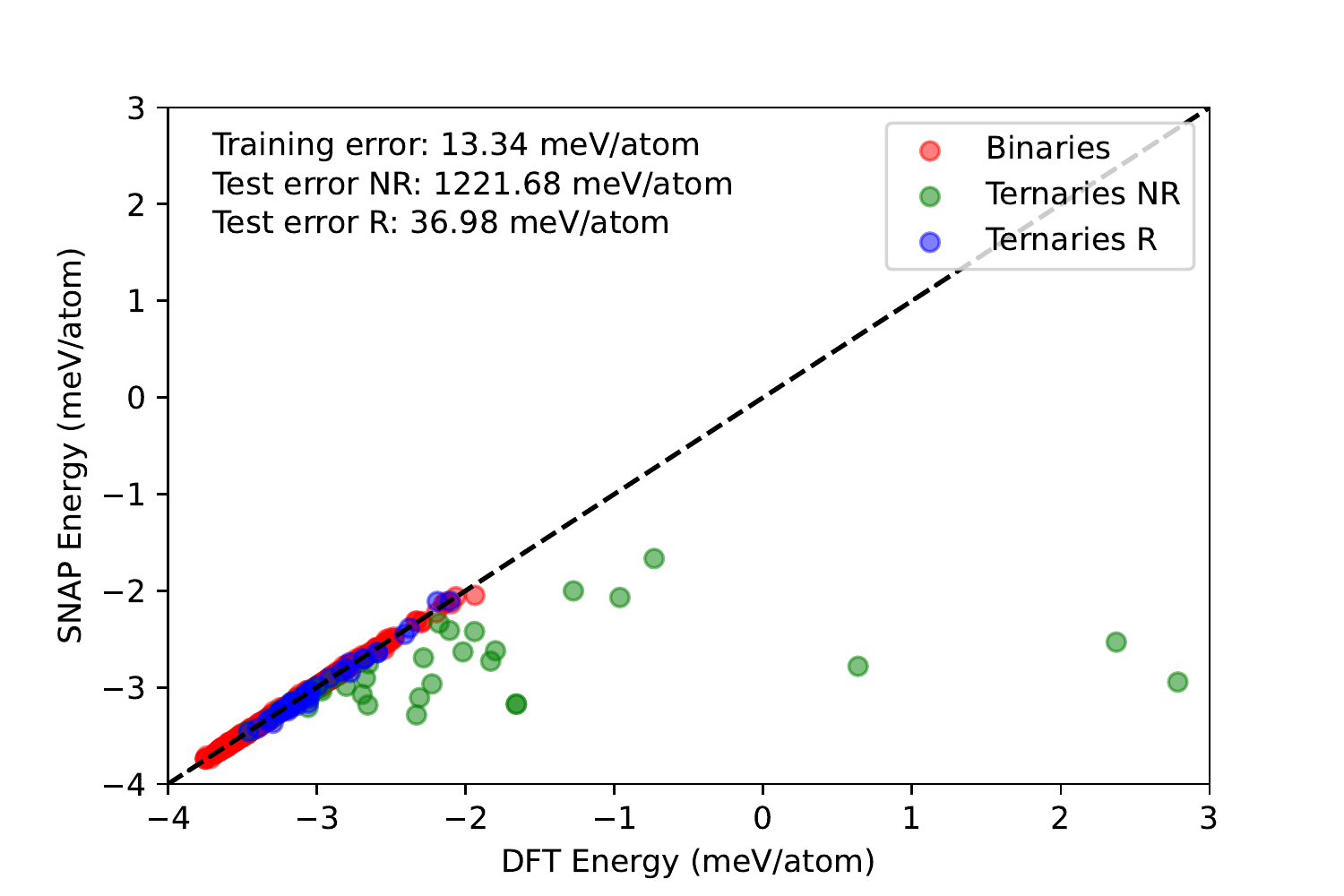}
    \caption{Parity plot, showing the SNAP-predicted energies against the DFT ones, for both the binary compounds 
    contained in AFLOWlib (training set) and for a new set of ternary phases either in their prototypes, non-relaxed (NR), 
    geometry or after full DFT relaxation (R). Here the SNAP models forming the ensemble have been trained on the entire 
    pool of binary phases as described in Section~\ref{section:TestingTernaries}. All data are for the Cu-Ag-Au system.}
    \label{fig:relaxtnonrelaxed_ternaries}
\end{figure}

In order to put our SNAP models against a more severe task, we now investigate whether these can be used to predict
the energy of novel prototypes not present in the AFLOWlib database. The dictionary method, as implemented in the AFLOW 
encyclopedia \cite{ HICKS2021110450,MEHL2017S1,HICKS2019S1}, is initially employed to create 42 new structures. 
These, in general, span a wide energy range and most of them are far away from the convex hull. Note, in fact, that the 
lattice parameters of these prototypes are not optimized, but just estimated through a Vegard-like law. Then, DFT relaxation 
is performed for all the newly created ternaries until the forces are below $10^{-3}$ eV/$\AA$. We finally assess the extrapolation 
ability of the SNAP models to predicting the ground state energy of both the relaxed (R) and the initial non-relaxed (NR) ternary 
prototypes. The ensemble SNAP used here is the same as the one introduced in Section \ref{section:TestingTernaries}. Our 
results are presented in Figure~\ref{fig:relaxtnonrelaxed_ternaries}.

The figure shows that once the structures are fully relaxed by DFT, the error remains satisfactory, confirming that the 
good energy estimate for ternary compounds does not remain limited to the AFLOWlib data. In this case, the RMSE grows 
from 17.46~meV/atom for the original ternary included in AFLOWlib to about 36.98~meV/atom for the new relaxed prototypes,
reflecting the more diverse set of structures generated for this test. In contrast, the error is significantly higher when one tries
to estimate the energy of the prototypes as constructed, namely before DFT relaxation. We now find a RMSE of 1.2~eV/atom,
with SNAP systematically underestimating the DFT energy. Interestingly, there is still a fraction of the created structures on which 
SNAP performs well, and these appear to have lattice parameters close to their relaxed ones. In general, this analysis proves that 
an ensemble SNAP model, constructed on binaries, can extrapolate to associated generic ternary structures when these are close 
to their equilibrium geometry. 

The remaining question is then whether or not the so-constructed SNAP can be used to drive the atomic relaxation. This is investigated by employing 
the LAMMPS package \cite{LAMMPS} for geometry minimisation, where the energy and force convergence criterion are 
set at $10^{-4}$ eV and $10^{-3}$ eV/\AA ~respectively. 
The relaxation is performed in two steps, where first the atomic positions are optimized, and then we relax both the cell parameters 
and the atomic positions. This procedure is repeated five times for each structure to ensure convergence. Again the 
mean SNAP model, averaged over all those of the ensemble, is chosen to perform the relaxation. We find that, although the resulting 
SNAP-optimized structures generally have a lower DFT-computed total energy than the unrelaxed ones, they are still far from the
optimal DFT-computed geometries. This means that, although the ensemble SNAP is capable of some relaxation, in general it is not able
to find the equilibrium structure. 

\begin{figure}
    \centering
    \includegraphics[width=9cm]{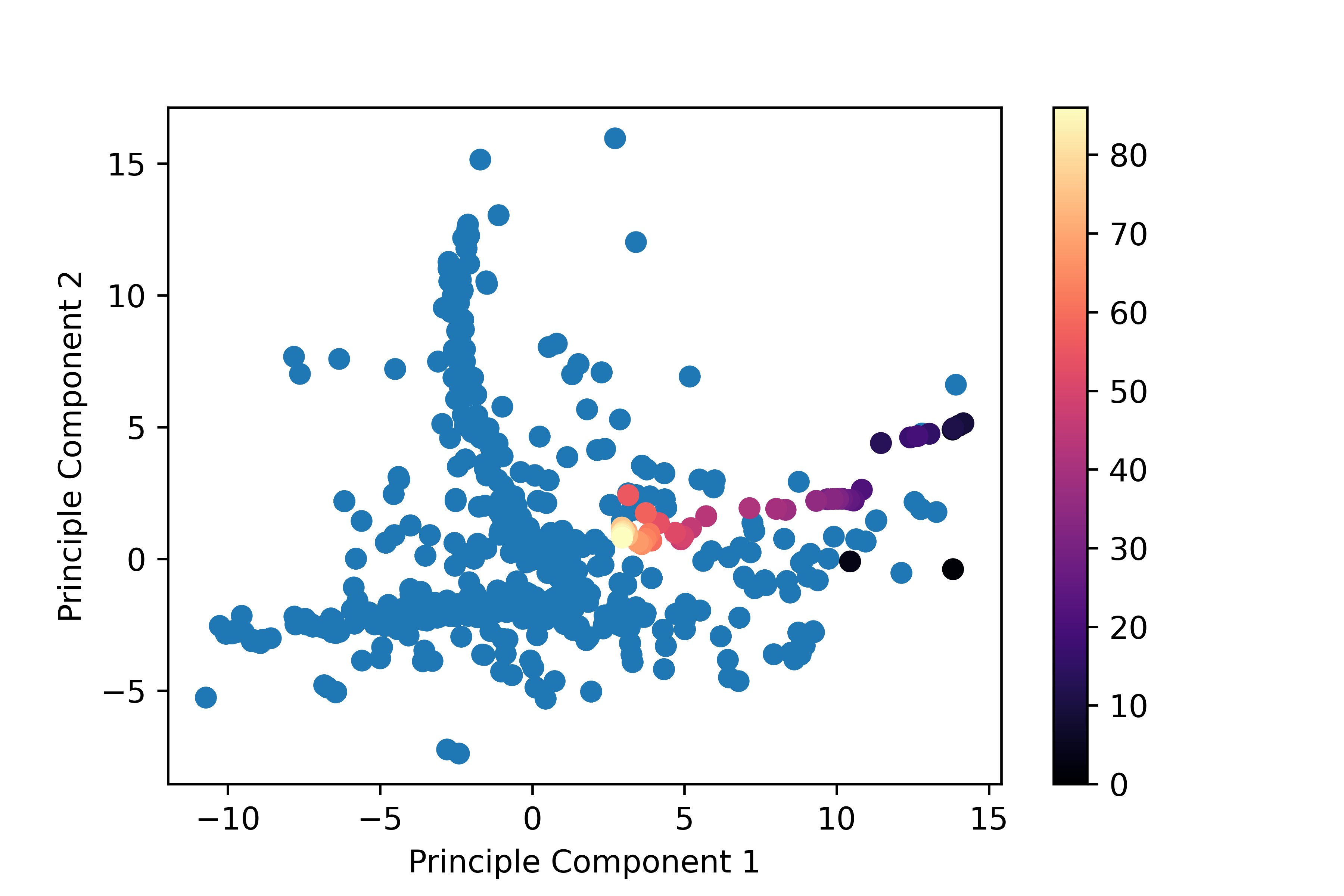}
    \caption{Plot of the first two principal components of a PCA performed over the Ag bispectrum of the 
    training set (blue circles). The coloured circles are for the ternary structures encountered during the \textit{ab initio} 
    relaxation path of a prototype ternary compound. The colour code encodes the relaxation-step index. Note that the
    DFT relaxation begins in a region poorly covered by the training set, but ends in a rather high-density region.}
    \label{fig:PCA}
\end{figure}

In order to obtain some insight into this aspect, principal component analysis (PCA) is performed on the feature vectors forming 
the training set and on those of the structures encountered along the DFT relaxation path of the ternary prototypes. This analysis 
is performed for each one of the three species considered here, Cu, Ag and Au. An illustrative plot of the PCA of the first two 
components is shown in Figure~\ref{fig:PCA} for Ag, and similar graphs have been obtained for the other two species. In the plot, 
blue circles represent the PCA components associated to structures included in the training set (binary phases), while the coloured 
ones are for a DFT relaxation trajectory starting from a new ternary prototype. Clearly, the binary feature space is quite heterogeneously 
distributed, with relatively large portions poorly known by SNAP. The DFT relaxation is found to travel sparsely populated regions, in 
particular for the starting structure. As a result, energy and forces during the SNAP-driven relaxation may be poorly predicted, 
since the relaxation trajectory has to travel regions in the parameter space of which the SNAP has little knowledge. This then results
into structures that differ from the \textit{ab initio} relaxed ones. 

There are a number of possible strategies to mitigate this problem. On the one hand, one can simply generate additional 
DFT-computed prototypes, cleverly covering a more uniform distribution of local atomic environments, and enlarge the 
SNAP training set. This avenue is usually pursued in conjunction with some active-learning strategy~\cite{bartok2010gaussian,shapeevactive,VASPotf,smith2021automated} 
to 
generate force fields for stable molecular dynamics simulations, but it often requires a rather large number of new DFT 
calculations. When the problem is that of constructing phase diagrams, this does not seem like an ideal solution, since 
the DFT effort is mostly spent in structures of no particular interest other than training the ML potential, and the total 
DFT-calculation count may be similar to that of computing the entire phase diagram from DFT alone. On the other hand, 
one can notice from figure~\ref{fig:PCA} that the typical DFT relaxation moves towards regions of the feature space well
covered by the binary training set. This means that constructing the initial prototypes with local atomic environments 
reflecting more closely those of the training set, may represent a better initial choice for novel ternary phases, that are 
closer to the final optimized structures and can be naturally relaxed with SNAP.

\subsection{Ternary phase diagrams for Ti-Mo-Pt and Cd-Hf-Rh}
In order to validate our results and to demonstrate that the method works well beyond noble metals, we have performed
the same workflow on two different new ternary systems, namely Ti-Mo-Pt and Cd-Hf-Rh. These have been selected 
based on two criteria: 1) they contain phases made of early (Ti and Hf), mid (Mo and Rh) and late (Pt and Cd) transition metals, 
thus presenting a chemical variety larger than that found in noble-metal compounds; 2) for these ternary systems AFLOWlib 
contains a high enough number of prototypes, enabling the training of a reliable model. In fact, these two systems contain more than 1,500 compounds, with around 90 ternaries for each set. It should 
be noted, however, that for each system, one binary combination is usually over-represented when compared to the 
others. These are Ti-Mo for Ti-Mo-Pt and Cd-Hf for Cd-Hf-Rh. 

\begin{figure}[!h]
    \centering
    \includegraphics[width=8cm]{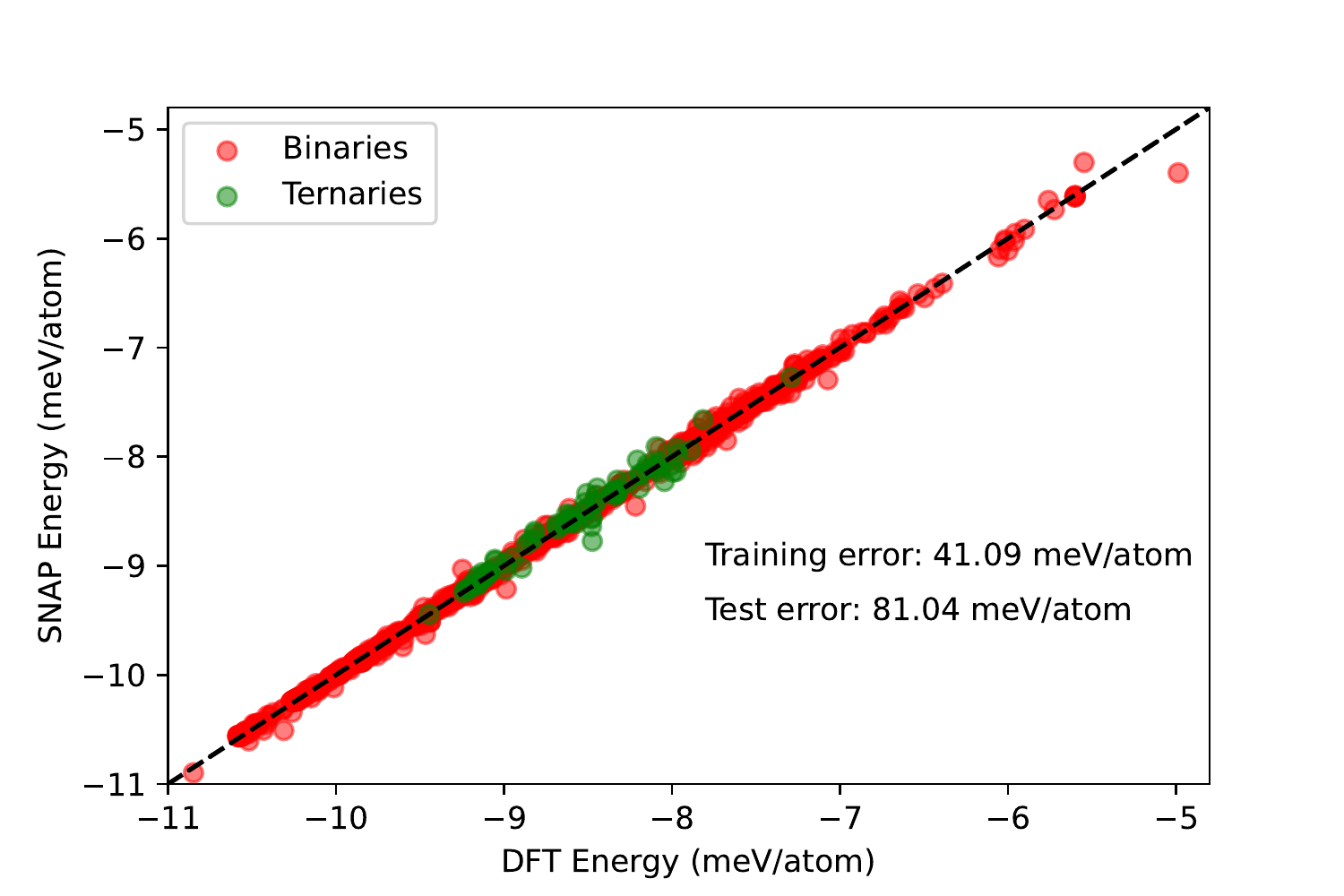}
    \includegraphics[width=8cm]{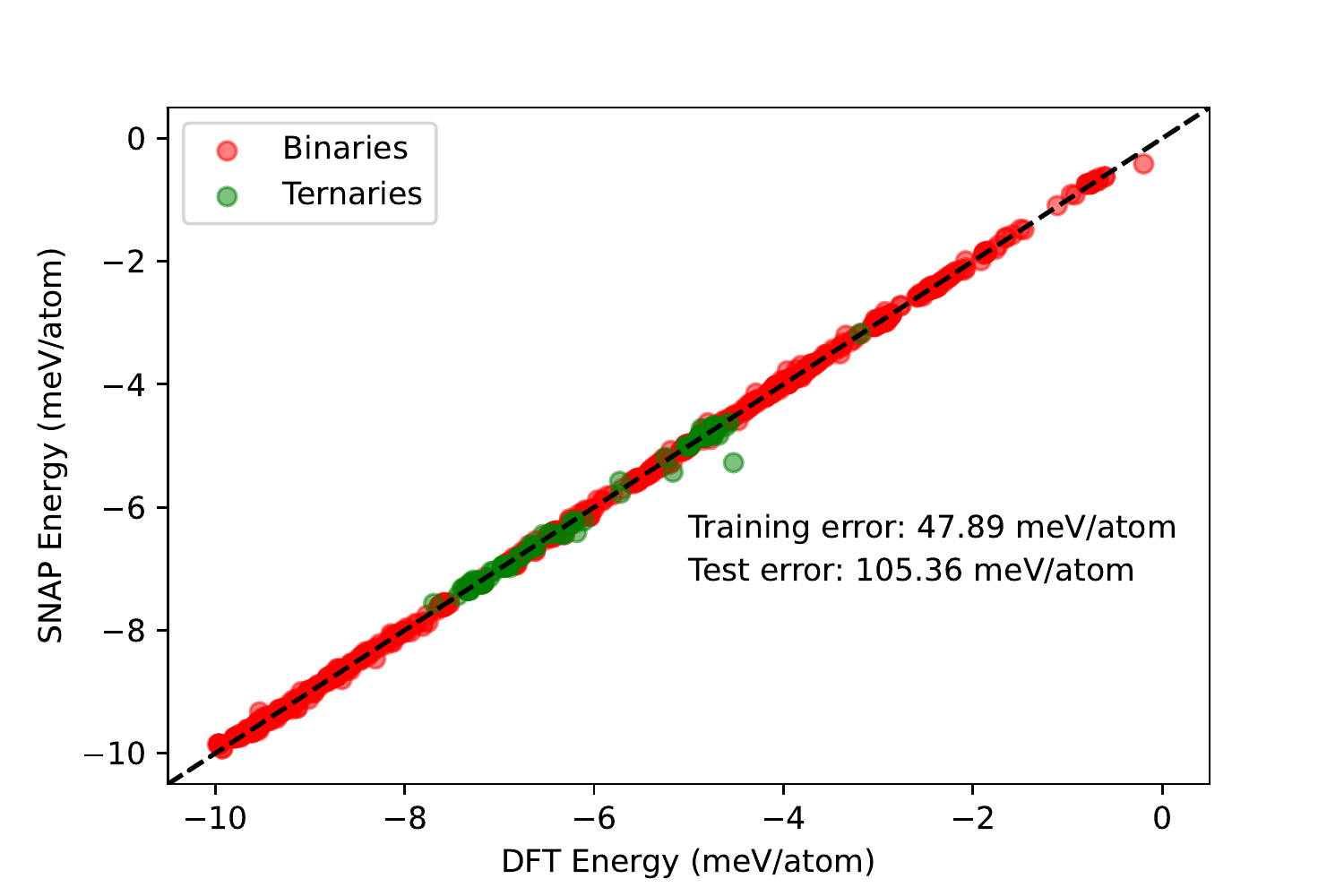}
    \caption{Parity plot, showing the SNAP-predicted energies against the DFT ones, for both the binary (training and 
    cross-validation set) and ternary (test set) compounds contained in AFLOWlib. Data are presented for the Ti-Mo-Pt
    (upper graph) and for the Cd-Hf-Rh (lower graph) ternary systems. Here the SNAP models forming the ensemble
    have been trained on the entire pool of binary phases. The RMSE is reported in the legend.}
    \label{fig:new_spec}
\end{figure}

The ensemble SNAP parity plots are shown in Figure \ref{fig:new_spec}, alongside with the RMSE of the fit. Also in this case
the training (cross-validation) set comprises the 80\% (20\%) of all available unary and binary compounds, the test set 
is only made of DFT-relaxed ternary phases, and carefully optimization of the hyperparameters, $R_{c}$, $J_\mathrm{max}$ 
and $w_{i}$, is performed. From the figures, it is clear that for these relaxed structures, taken from AFLOWlib's convex hull, 
the predictions are satisfactory and the parity line is followed closely. The absolute errors on all sets are higher than those found for 
Ag-Au-Cu, but the total-energy percentage error is similar. In fact, for Ag-Au-Cu, the percentage error relative to the average 
energy per atom was of 0.42\% for the training set and 0.53\% for the test set. These must be compared to 0.46\% and 0.94\%
for Mo-Pt-Ti and to 0.80\% and 1.69\% for Cd-Hf-Rh. It should be noted that the ternary convex hull diagrams for Mo-Pt-Ti and 
Cd-Hf-Rh are much deeper that that of Cu-Ag-Au as the lowest points are at an enthalpy of formation of 951~meV/atom and 
920.54~meV/atom respectively, compared to 61~meV/atom for Ag-Au-Cu. This makes the SNAP absolute error less significant. 

\subsection{Error Metrics}

The aim of this work is to develop a ML model surrogate to DFT to predict total energies and hence to identify stable ternary 
compounds. As shown in the previous sections, a SNAP potential constructed over binary phases is able to provide
accurate energy predictions for ternary compounds only close to their equilibrium structure, but it is not reliable to perform
relaxation. This is because during the relaxation process the system cross regions in the parameter space poorly 
covered by the training set. Therefore, it would be of interest to have a metric capable of distinguishing structures that are 
far from the feature space spanned by the training set from those that are within it. We expect SNAP to perform well for the 
second set of structures but not for the first. In this section, three such metrics are presented and their effectiveness is assessed. 

The first error metric used is the Euclidean distance, $d_\mathrm{min}$, which can be here defined in different ways. The 
easiest approach is to take the distance between the feature vector of the test system and the average feature vector of 
the training set. Unfortunately, the total feature vector is made up of a sum of bispectrum components, some of which are 
associated to different atomic species. Instead, the so-defined distance is evaluated independently of each species. For a given 
species, two sets are created, the set of all bispectrum components in the training set and all those in the test set. Then the 
Euclidean distance between the vectors of each set is evaluated, and the metric used is the minimum of all these distances, 
namely
\begin{equation}
    d_\mathrm{min} = \textrm{min} \left \{ ||\vec{B_{i}} - \vec{B_{j}}||\right \}_{i\in \Omega_\mathrm{Training},\:\: j\in \Omega_\mathrm{Test}}\:,
\end{equation}
where $\Omega_\mathrm{Training}$ ($\Omega_\mathrm{Test}$) is the ensemble of structures contained in the training (test) set. 
Since a given compound could have several atoms of the same species, the maximum of all $d_\mathrm{min}$ is assigned to that 
compound. Note that for each test ternary system, three values of this distance metric are obtained, one for each species.

The second error metric tested is the extrapolation grade, $\gamma$, as introduced by Podryabinkin \textit{et al.} \cite{shapeevactive}. 
The approach followed here is closest to their Query Strategy 3, as it is specie-wise defined. From the bispectrum components of all the $k$ configurations in the training set, a rectangular $k\times m$ matrix, $\mathbf{B}$, is formed, where $m$ is the length of the feature vector. The {\sc maxvol}~\cite{maxvol}
algorithm 
is then applied to find the $m\times m$ sub-matrix $\mathbf{A}$ with maximal determinant.
This corresponds to the selection of the active set. Then, for a new structure outside of the training set, the extrapolation 
coefficient $\gamma$ is calculated for each bispectrum component of that system and the maximum value is selected. As for the distance 
metric, this is done species-wise. 

The final error metric chosen is the standard deviation, $\sigma$, of the SNAPs prediction. As described above, an ensemble of SNAP 
models is trained over different batches of the training set. The mean value of the predictions of each model is taken as the prediction 
of the ensemble. This approach also allows one to obtain the standard deviation of the predictions, which is used here as an error metric.

\begin{figure}[!h]
    \centering
    \includegraphics[width=8cm]{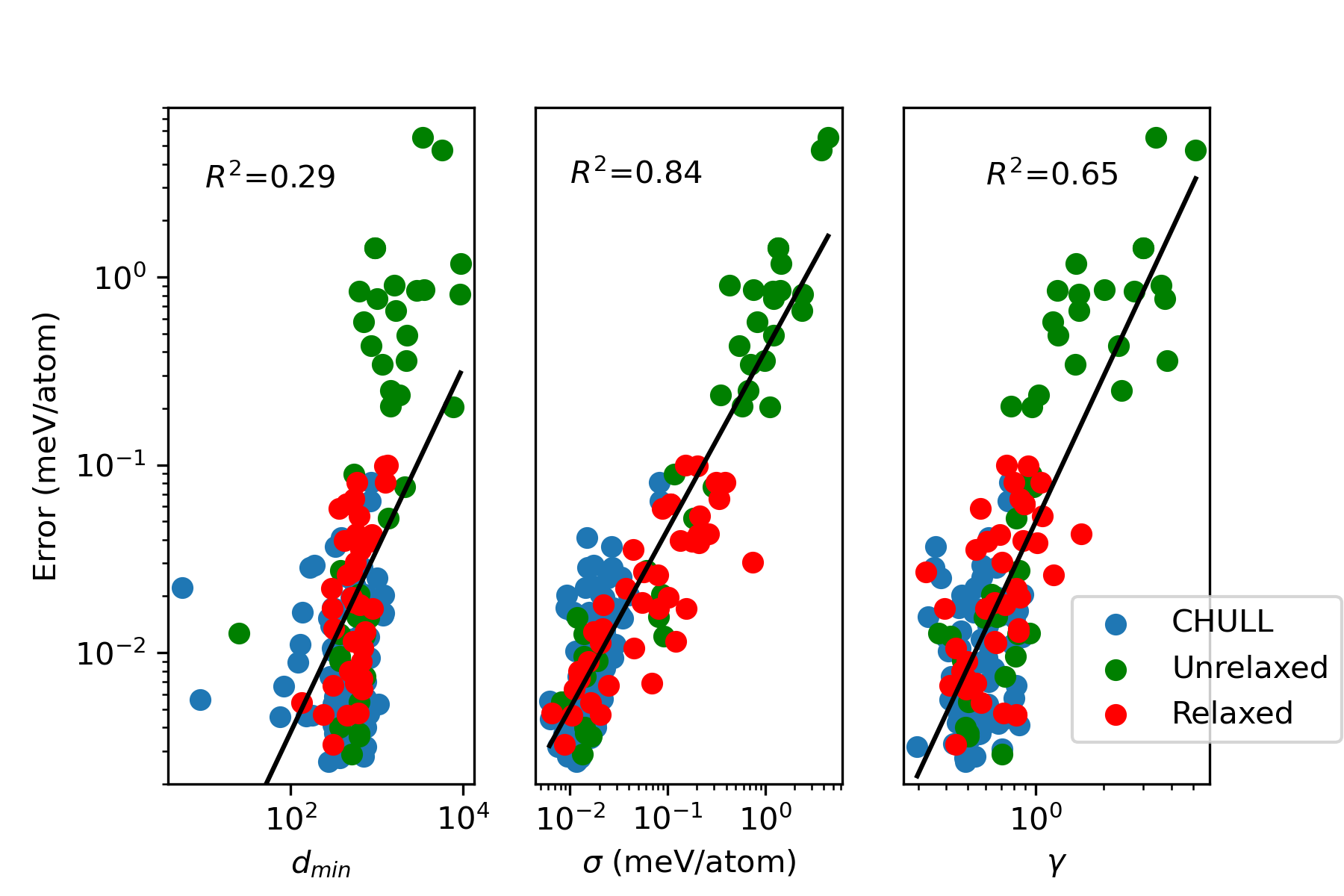}
    \caption{Plot of the SNAP error against our defined three different metrics for the three ternary datasets in the Cu-Ag-Au system: the 
    compounds residing on the convex hull diagram (CHULL), the unrelaxed new prototypes (Unrelaxed) and the new prototypes after relaxation (Relaxed). The metric on 
    the $x$-axis from left to right is: the distance from the training $d_\mathrm{min}$, the standard deviation, $\sigma$, and the extrapolation 
    grade, $\gamma$. These are shown for Ag for $d_\mathrm{min}$ and for Cu for $\gamma$ (note that $\sigma$ is species independent). 
    The black lines show the linear fits on the data points, and the associated $R^{2}$ is given.}
    \label{fig:error_metrics}
\end{figure}

These metrics are assessed and compared by evaluating their correlation with the SNAP absolute error. A variety of three test sets in the 
Cu-Ag-Au ternary system is used for this exercise, namely the ternaries presented in Section \ref{section:TestingTernaries} (compounds 
on the convex hull), as well as the unrelaxed and relaxed prototype ternary compounds introduced in Section \ref{sec:TernaryProto}. 
In figure~\ref{fig:error_metrics} we present the SNAP mean absolute error against the error metric for each phase contained within 
the sets. These are plotted on a log scale, since both the error metrics and errors span several orders of magnitude. A linear fitting is 
performed on all data points for each plot and the associated values of $R^{2}$ are also given. For the $d_\mathrm{min}$ and the $\gamma$ 
metrics, the results are shown for the bispectrum of a specific specie, Ag and Cu respectively. As discussed previously, the structures in 
the AFLOWlib CHULL and in the relaxed datasets, that are close to equilibrium, have lower errors in general, with a range 78-97~meV/atom, 
whereas it is $\sim$5~eV/atom for the structures of the unrelaxed dataset. This implies that as a general feature of all plots, there is a group of points 
at lower errors that are within a small range of the error metrics, namely there is little correlation between metric and error. This is to be expected, 
since this range is of the same order of magnitude as the cross-validation error of the SNAP model. The metrics then become more relevant 
at larger errors, where the structures are expected to be outside of the training range, as for those in the unrelaxed dataset. 

Going into more detail, we find that the $d_\mathrm{min}$ minimum distance metric is not much larger for the structures with large errors, 
resulting in a low $R^{2}$. This indicates that the Euclidean distance between test and training vectors is not adequate for measuring the 
extent of extrapolation in our work. In contrast, the extrapolation grade, $\gamma$, and especially the standard deviation of the model 
prediction, $\sigma$, correlate far better with the error, with $R^{2}$ values of 0.65 and 0.84 respectively. Both of these metrics are significantly 
larger for the structures of the unrelaxed datasets compared to those in the AFLOWlib CHULL and relaxed datasets. It is interesting to note 
that the best performing metric is the standard deviation metric, which is not a specie-defined metric, so it can only be calculated for an entire 
structure. The minimum distance metric and extrapolation grade can be made global for one structure by averaging over the metrics of the individual 
atoms. This increases the $R^{2}$ values to 0.51 and 0.74 respectively. The standard deviation is also the only metric that makes use of the 
performance of the trained model, since it relies on the SNAP coefficients. The other two metrics only make use of the feature vectors of the training 
set.

\section{Conclusions}
In conclusions, we have examined the extrapolation ability of SNAP models trained on data extracted from binary phase diagrams to predict the
ground-state energy and hence the thermodynamical stability of unseen ternary prototypes. Such analysis has been performed over a noble-metal 
ternary system, Cu-Ag-Au, and over two mixed systems, namely Ti-Mo-Pt and Cd-Hf-Rh. Furthermore, we have explored different error metrics 
in the search for a criterion to detect structures that are far from the training set. In this way one could easily identify prototypes for which 
SNAP would struggle with.

We have been able to establish that SNAP models trained on binary-phase data are able to predict the stability of novel ternaries, if those are 
close to their energy minimum, namely if they are relaxed. In contrast, the predictions start to worsen as the structures move away from equilibrium. 
This drawback could be fixed by an active-learning step, which introduces unrelaxed binary structures to the model. Such models' ability to predict 
the ground-state energy of unseen prototypes can be used as first level screening components in prototype creation algorithms that explore large 
materials spaces, where high-throughput is more important than high-accuracy.  

Different error metrics have been investigated in order to find the most suitable to distinguish structures that the training set contains no 
information about. We have found that the best performing error metric is the ensemble SNAP standard deviation, $\sigma$. This is a global 
metric, which is independent from the particular specie, and it is simple and efficient to calculate. Based on this results, we can conclude that 
the integration of machine-learning steps in the construction of ternary phase diagrams is indeed possible, once a clever strategy for the 
generation of the prototypes is identified. 

\begin{acknowledgments}
This work has been supported by the Irish Research Council Advanced Laureate Award (IRCLA/2019/127),
and by the Irish Research Council postgraduate program (MC). We acknowledge the DJEI/DES/SFI/HEA Irish 
Centre for High-End Computing (ICHEC) and Trinity Centre for High Performance Computing (TCHPC) for the 
provision of computational resources. 
\end{acknowledgments}

\section*{Author Contributions}

This section is written according to the CRediT system.
MM and HR contributed equally to this work. MM and HR contributed to conceptualisation, methodology, software, data curation, formal analysis, investigation, validation, visualisation, writing the original draft, as well as reviewing and editing the manuscript. MC contributed to investigation, formal analysis, software as well as reviewing and editing the manuscript. SS contributed to conceptualisation, funding acquisition, project administration, resources, supervision, as well as reviewing and editing the manuscript.

\section{Appendix}

\subsection{Learning Curves}
In this subsection the learning curves for the models trained on the three binary systems of Cu-Ag-Au are presented. These are used to 
establish the amount of data needed to train the SNAP model.

\begin{figure}
    \centering
    \includegraphics[width=8cm]{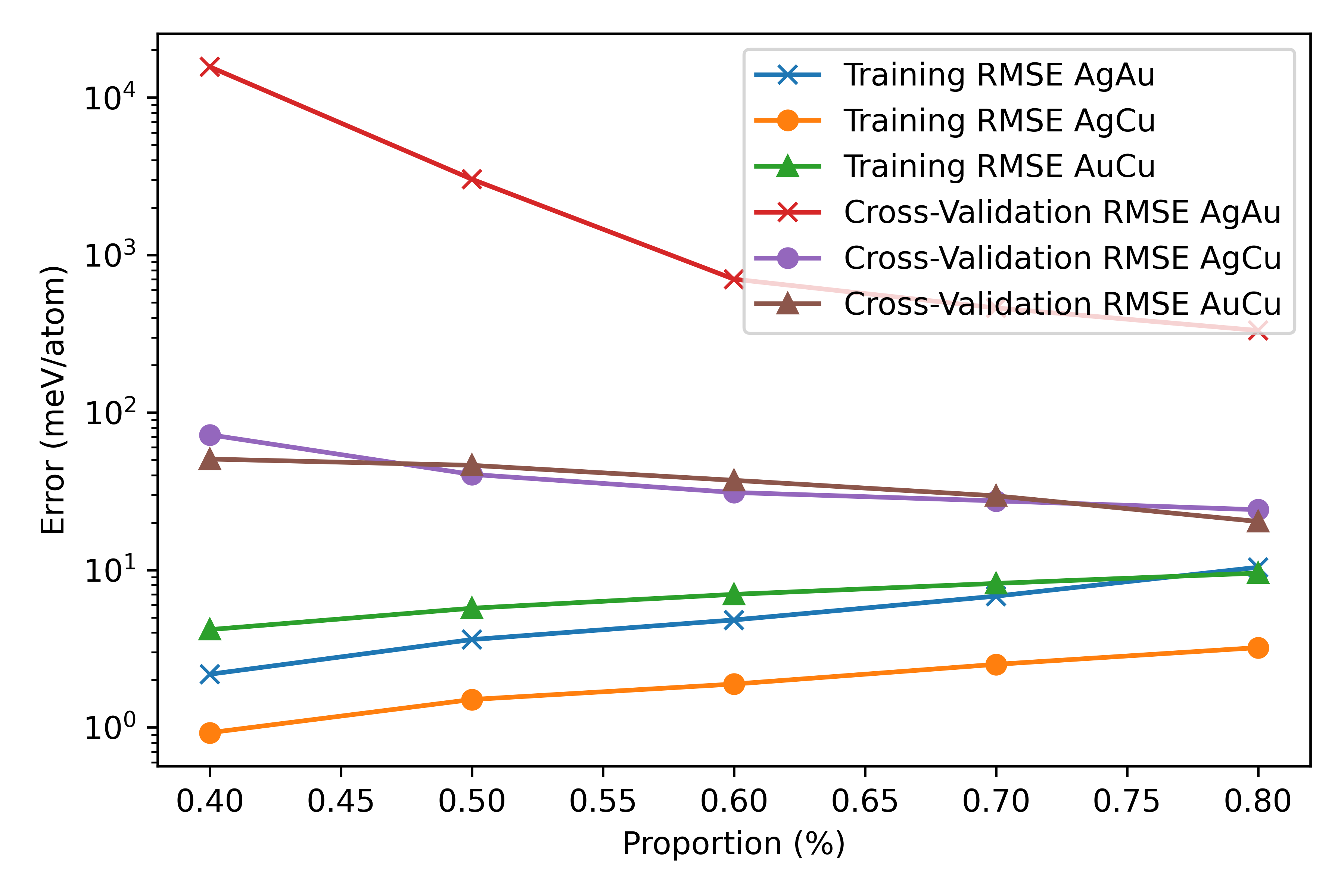}
    \caption{Learning curves of the models trained on the binary data. As one can notice the optimal value for the data splitting is at 80\% for the 
    training sets and 20\% for the test sets.}
    \label{fig:LC_CuAgAu}
\end{figure}
As it can be seen from Fig.~\ref{fig:LC_CuAgAu}, a 80\% (training)/20\% (cross validation) split appears ideal. Note that similar
curves are obtained for the other two ternary systems investigated (not presented here). 

\subsection{Hyperparameter Optimisation}

Following a grid search for the hyperparameter optimisation, the values given in Table \ref{tab:HP_single_binaries} are found to 
minimise the RMSE for the single binary SNAP models.

\begin{table}
\centering
\caption{Summary of the optimal hyperparameters for the SNAP models individually trained over the binary systems. 
Here $w_{X}$ and $w_{Y}$ refer to the weights of the X and Y species of the X-Y binary system.}
\label{tab:HP_single_binaries}
\begin{tabular}{l||l|l|l}
\hline\hline
Hyperparameter	& Ag-Au & Ag-Cu & Au-Cu \\
\hline
$J_\mathrm{max}$           & 3	& 3  & 3\\
$R_{c}$          & 3.2	& 4.2  & 4.6\\
$w_{\mathrm{X}}$	& 1.0	& 0.6  & -0.8\\
$w_{\mathrm{Y}}$	& 0.8	& 0.4 & -0.4\\
\hline\hline
\end{tabular}
\end{table}
\vspace{1cm}
The same procedure was followed to minimise the RMSE for the SNAP models trained on all three binaries. The results are given in 
Table \ref{tab:HP_all_binaries}.
\begin{table}
\centering
\caption{Summary of the hyperparameters used for the three binaries SNAP models.}
\label{tab:HP_all_binaries}
\begin{tabular}{l||l}
\hline\hline
Hyperparameter	& Ag-Au-Cu \\
\hline
$J_{max}$           & 4	\\
$R_{c}$          & 4.6\\
$w_\mathrm{Ag}$	& -3.0\\
$w_\mathrm{Au}$	& -4.0\\
$w_\mathrm{Cu}$	& -3.0\\
\hline\hline
\end{tabular}
\end{table}

\vspace{10cm}
\pagebreak

\bibliography{tex-new.bib}

\end{document}